# One-step synthesis of van der Waals heterostructures of graphene and 2D superconducting α-Mo$_2$C

Jia-Bin Qiao[1,§], Yue Gong[2,§], Wei-Jie Zuo[1], Yi-Cong Wei[1], Dong-Lin Ma[1], Hong Yang[1],

Ning Yang[1], Kai-Yao Qiao[1], Jin-An Shi[2], Lin Gu[2,3,4] and Lin He[1,*]

[1]Center for Advanced Quantum Studies, Department of Physics, Beijing Normal University, Beijing, 100875, People's Republic of China
[2]Beijing National Laboratory for Condensed Matter Physics, Institute of Physics, Chinese Academy of Sciences, Beijing 100190, China
[3]Collaborative Innovation Center of Quantum Matter, Beijing 100190, China
[4]School of Physical Sciences, University of Chinese Academy of Sciences, Beijing 100190, China
[§]These authors contributed equally to this work.
*Correspondence and requests for materials should be addressed to L.H. (E-mail: helin@bnu.edu.cn).

**Assembling different two-dimensional (2D) crystals, covering a very broad range of properties, into van der Waals (vdW) heterostructures enables the unprecedented possibilities for combining the best of different ingredients in one objective material. So far, metallic, semiconducting, and insulating 2D crystals have been used successfully in making functional vdW heterostructures with properties by design. Here, we expand 2D superconducting crystals as a building block of the vdW hererostructures. A one-step growth of large-scale high-quality vdW heterostructures of graphene and 2D superconducting α-Mo$_2$C by using chemical vapor deposition (CVD) method is reported. The superconductivity and its 2D nature of the heterostructures are characterized by our scanning tunneling microscopy (STM) measurements. This adds the 2D superconductivity, the most attractive property of condensed matter physics, to the vdW heterostructures.**

Recently, van der Waals (vdW) heterostructures obtained via layer-by-layer stacking of different 2D crystals[1-4] have attracted great interests[5-15]. The 2D crystals with various properties are resembling atomic-scale Lego blocks in building the ultimate vdW heterostructures. Interestingly, such artificial materials engineered with atomic accuracy exhibit properties distinct from their components and, sometimes, they even show exotic phenomena that beyond one could have imaged[8-19]. For example, by simply stacking graphene on top of an insulating hexagonal boron nitride (hBN) results in the emergence of topological currents in such a system[16,17]. Now, it is fair to say that scientists have already achieved great success by using 2D vdW heterostructures to realize materials with a variety of properties. However, 2D superconductivity, as the most attractive property of condensed matter physics, is still not included yet. Very recently, the newly discovered 2D superconductors[20-26] provide unprecedented opportunities to bring the 2D superconductivity into vdW heterostructures. Via proximity effect, the superconducting phase of superconductors can be introduced into non-superconducting crystals[27-32]. Therefore, it is reasonable to expect that the combination of the 2D superconductors and other 2D crystals can lead to many extraordinary physical properties.

Here we address this issue for the first time and report a feasible one-step growth of large-scale high-quality vdW heterostructures of graphene and 2D superconducting α-$Mo_2C$. Previously, graphene was grown successfully on groups IVB-VIB metal foils (including Mo foils) by chemical vapor deposition (CVD) method[33-35]. In this work, we add a Cu foil on top of a Mo foil to direct grow the graphene/α-$Mo_2C$ heterostructures.

Above 1085 ℃ (the melting point of Cu), the newly added Cu foil not only governs the diffusion process of Mo atoms to tune the chemical reaction rate, but also manipulates the growth mechanisms, thus realizing one-step growth of well-defined vdW heterostructures of graphene and α-$Mo_2C$.

**Results**

**Synthesis and characterization of the graphene/α-$Mo_2C$ vdW heterostructures.**
Figure 1a shows schematic drawings of the growth mechanism for the graphene/α-$Mo_2C$ heterostructures. Above 1085 ℃, Mo atoms diffuse into the melted Cu foil to form Cu-Mo alloy. Then, we can control the final products by tuning the ratio, $\lambda$, of methane ($CH_4$) to hydrogen ($H_2$) during the growth process. At low ratio ($\lambda \leqslant$ 1:500), we obtained well-defined 2D α-$Mo_2C$ crystals due to the predominant "carbide growth" mechanism attributing to the large bonding strength between Mo and C atoms[23]. However, at relatively high ratio ($\lambda >$ 1:500), there is a high level of C source and the onset of "graphene growth" process is triggered. The newly added Cu foil plays a vital role in the formation of the vdW heterostructures. It competes with the Mo component to capture C atoms for surface-catalytic graphene growth[36]. Assisted with the graphitic catalysis of pre-formed α-$Mo_2C$ crystals, large-scale high-quality graphene grows on both the exposed Cu (Cu-Mo alloy) surface and the α-$Mo_2C$ crystals, thus generating the 2D superconducting vdW heterostructures of graphene and α-$Mo_2C$, as shown in Fig. 1b. The heterostructures usually show regular shapes, such as triangle, rectangle and hexagon, tailored by the underlying α-$Mo_2C$ sheets (see Fig. 1b-f and Supplementary Information for more details). Importantly, it is convenient to control the dimension and thickness of the α-Mo$_2$C sheets in the heterostructures by changing the growth parameters, such as the ratio $\lambda$, the growth temperature and time, as well as the thickness of the Cu foil (see Supplementary Information for more details on the size adjustment). In our experiment, the size of the α-Mo$_2$C sheets in the heterostructures can be changed from a few micrometers to over 50 μm, and the thickness can be varied from a few nanometers to hundreds of nanometers.

Figure 1g shows a representative Raman spectrum of the obtained vdW heterostructures of graphene and α-Mo$_2$C transferred on a SiO$_2$/Si substrate (see Methods for more details). The three peaks located at low wavenumbers are characteristic signals of the α-Mo$_2$C[37] and the G and 2D peaks of graphene[38] are also clearly observed in the Raman spectrum. In our experiment, we find that the 2D peak exhibits a symmetric single Lorentzian line shape and the intensity ratio of G to 2D peaks falls into a range of 0.4 to 0.6. These features are the hallmarks of Raman spectra of single-layer graphene, in accord with the favorable growth of monolayer graphene on TMCs[33]. Compared with the spectrum of surrounding graphene sheet, a pronounced D peak emerges in the spectrum of graphene in the heterostructure (Fig. 1g), which may originate from the $sp^3$-hybridized defects caused by the strong bonding between the Mo and C atoms. The vdW heterostructures were further characterized by X-ray diffraction (XRD), energy dispersive X-ray spectra (EDS), atomic force microscopy (AFM) and X-ray photoelectron spectra (XPS) (see Supplementary Fig. S10 to S13 for more details). All these measurements demonstrated the creation of vdW heterostructures of graphene and 2D α-Mo$_2$C crystals.

The atomic structure of the 2D vdW heterostructures has been investigated by high-resolution scanning transmission electron microscopy (HR-STEM), as visualized in Fig. 2. Figure 2a shows a low-magnification high-angle annular dark-field (HAADF)-STEM image of a hexagonal heterostructure. Figure 2b shows selective area electron diffraction (SAED) pattern along the [100] zone axis of the heterostructure. The coexistence of two types of ED patterns from the α-$Mo_2C$ and monolayer graphene confirms the formation of vdW heterostructures (see Supplementary Fig. S16 for more details). According to the spatial distribution of the two sets of spots, the rotation orientation between α-$Mo_2C$ (100) and graphene lattices is extracted as ~ 11 °. Fig. 2c shows the atomic-resolution HAADF-STEM image of the heterostructure. Despite the absence of graphene lattice due to the insensibility on the light C atoms of single-layer graphene sheet, the well-defined α-$Mo_2C$ (100) lattices in an orthorhombic arrangement are clearly revealed, where six Mo atoms are arranged to a hexagonal close packed (hcp) structure and a C atom is located at the center of the hcp structure. Moreover, the HAADF-STEM cross-section image, as shown in Fig. 2d, unveils the ABAB-stacking of Mo layers, which is consistent well with the Mo lattice in α-$Mo_2C$ (010) surface. Obviously, all the STEM measurements validate the realization of the graphene/α-$Mo_2C$ heterostructures (see Supplementary Fig. S13 and S15 for more details).

**Topography and spectroscopy studies of the graphene/α-$Mo_2C$ vdW heterostructures.** The well-defined graphene/α-$Mo_2C$ vdW heterostructures provide an unprecedented platform to explore the proximity-induced 2D superconductivity by using scanning tunneling microscopy/spectroscopy (STM/STS)[32,39]. Figure 3a shows a typical STM image of a monolayer graphene on the α-Mo$_2$C crystal (i.e., a graphene/α-Mo$_2$C heterostructure), which is characterized by peculiar patterns (Moiré patterns) that are generated by the lattice mismatch between surface of α-Mo$_2$C and graphene. The observed features differ from the surface topography of graphene grown on Cu substrate (see Supplementary Fig. S17 and S18). Figure 3b shows an enlarged STM image of the heterostructure surface, where hexagonal honeycomb lattice of monolayer graphene overlapping the Moiré patterns is clearly observed. The coexistence of graphene and α-Mo$_2$C in the heterostructure is further confirmed by the fast Fourier transform (FFT) of the atomic-resolution STM image, as shown in Fig. 3c. According to the FFT image, the rotated angle between graphene and α-Mo$_2$C lattices in the heterostructure is about $(14 \pm 2)°$, which is nearly equal to the results extracted from SAED patterns. It is noticeable that there are six well-defined Bragg spots of graphene reciprocal lattice, whereas only one pair of the spots corresponding to α-Mo$_2$C (100) reciprocal lattice are observed in the FFT of the STM image. This may possibly originate from the STM measurements as the STM probes predominantly the top graphene layer.

Fig. 3d shows three representative tunneling spectra of the vdW heterostructure recorded at different temperatures. All the spectra exhibit V-shaped curves, as expected to be observed for the stable heterostructure where graphene monolayer is supported by metallic α-Mo$_2$C surface[40]. The Dirac point of graphene in the heterostructure is estimated to be at about -300 meV (see Fig. 3d). Zooming in the spectrum measured at 400 mK, a superconducting gap appears around $E_F$, as shown in Fig. 3e. The superconducting gap in the spectrum can be well reproduced by BCS Dynes fit[41] (see Supplementary Information for more details), direct revealing the proximity-induced superconductivity in graphene of the heterostructure.

**Temperature and field dependence of the proximity-induced superconductivity.** To further study the 2D superconductivity of the heterostructures, we carried out spectroscopic measurements at various temperature and in the presence of perpendicular magnetic fields. In Fig. 4a, we show the temperature evolution of the tunneling spectra measured at the same position of the heterostructure. The coherence peaks associating with the gaps are suppressed with increasing temperature, and finally vanish above $T \sim 3.9$ K. The superconducting gap of the heterostructure $\Delta$, which is obtained by Dynes fitting of the spectra, as a function of temperature is shown in Fig. 4b. The solid curve in Fig. 4b represents the best fit to the data by the BCS gap function. Based on the fitting, we extract the superconducting gap $\Delta(0)$ and $T_c$ as 1.58 meV and 4.0 K, respectively, leading to a gap ratio $2\Delta(0)/k_BT_c$ of $\sim 9.1$. The ratio value is much larger than the one $2\Delta(0)/k_BT_c \sim 3.53$ predicted by the BCS theory in the weak coupling regime[42], indicating strong coupling mechanism in the system.

We also carried out spectroscopic measurements of the heterostructure in perpendicular magnetic fields. As shown in Fig. 4c, the superconducting gap decreases with increasing the magnetic fields and disappear above $\sim 200$ mT at 0.4 K. Such a result well matches the predicted relationship between the gap and the applied fields[43], $\Delta(H)/\Delta(0) = \sqrt{1-H/H_{c2}}$.

**Discussion**

Furthermore, the Ginzburg-Landau (GL) coherence length, $\xi_{GL}(0)$, is extracted as 38.5 nm from the general linearized GL relation, $H_{c2}(T) = \Phi_0/2\pi\xi_{GL}^2(0)(1-T/T_c)$ ($\Phi_0$ is the magnetic flux quantum). The value is larger than the thickness $d_{SC}$ of the studied α-Mo$_2$C sheets, which are in the range of 10 ~ 30 nm (see Supplementary Fig. S7). This demonstrates the 2D superconducting behavior of the studied heterostructure. In our experiment, we found that the superconducting transition temperature $T_c$ and the critical field $\mu_0 H_{c2}$ (T) of the heterostructures can be tuned by simply varying the thickness of the α-Mo$_2$C sheets in the heterostructures, as shown in Fig. 4e. Because the thickness of the α-Mo$_2$C sheets is easy to control in the growth process, therefore, this provides a facile route to tune the superconductivity of the vdW heterostructures.

Our experiments have turned on the access to one-step synthesis of large-scale high-quality vdW heterostructures of graphene and 2D superconducting α-Mo$_2$C. The well-defined superconducting heterostructures with flexible size adjustment not only offer an ideal system to further study the 2D superconducting properties of the artificial materials, but also open up the possibility to create designable high-quality superconducting heterostructure devices.

## Methods

### One-step growth and transfer of graphene/α-Mo$_2$C heterostructures.

A Cu foil (10 μm thick, 99.8% purity; 12.5 μm thick, 99.9% purity; 25 μm thick, 99.8% purity, Alfa Aesar) is supported by a Mo foil (0.1 mm thick, 99.95% purity, Alfa Aesar) and then together they were loaded into a horizontal tube furnace (Xiamen G-CVD system). The substrate was heated above 1085 ℃ under $H_2$ (500 sccm). Subsequently, gaseous mixtures of $CH_4$ and $H_2$ with different proportions (more than 1:500) were introduced into the reaction tube for the fabrication of graphene/α-Mo2C heterostructures at ambient pressure for different durations (5 ~ 60 mins). Finally the system was rapidly cooled down with $H_2$ (100 sccm).

Wet etching of the Cu foil was performed to transfer the heterostructures onto the $SiO_2$/Si substrates and TEM grids. The surface of the heterostructures was spin-coated by a thin layer of polymathic methacrylate (PMMA) and then cured at 150 ℃ for 30 mins. After that, the PMMA-coated specimens were etched by a 0.2M $(NH_4)_2S_2O_8$ solution, resulting in a free-standing PMMA/heterostructures film floating on the surface of the solution. The film was thoroughly washed with deionized water and then transferred to the target substrates. The PMMA film was removed by low-pressure thermal annealing with $H_2$ : Ar = 20 : 50 at ~ 350℃, and then we obtain clean and high-quality heterostructures.

**Characterization of graphene/α-Mo$_2$C heterostructures.**

*Optical measurements*. Optical images of the as-grown and transferred heterostructures on Cu/Mo substrates were taken by an Olympus microscope (Olympus BX51M). Raman spectra were taken by a Raman spectroscopy system *Nanofinder 30* with a laser excitation wavelength of 532 nm and power of 20.7 mW.

*XRD, SEM, EDS, XPS, AFM measurements*. XRD spectra were taken by Shimadzu LabX XRD- 6000 with Cu *Kα* radiation. SEM images with EDS measurements were taken by Hitachi S4800 FESEM system. XPS measurements were taken by ESCALAB 250Xi system. AFM measurements were performed on AFM MultiMode 8 system.

*TEM measurements*. The cross section sample was prepared using focused ion beam milling, FEI Helios 600i. The STEM images were performed using a probe aberration corrected JEM-ARM200F microscope equipped with cold emission gun. The electron diffraction patterns were performed in a Philips CM200 microscope. EDS mapping was acquired using FEI Tecnai F20 microscope. All TEM data were operated at 200kV.

*STM/STS measurements*. The 77 K, 4.2 K and 0.4 K (including the variable temperature/magnetic field experiments) measurements were performed on the ultrahigh vacuum single-probe scanning probe microscopes USM-1400, USM-1500 and USM-1300 (with the magnetic field up to 15T) from UNISOKU, respectively. All the STM and STS measurements were performed in the untrahigh vacuum chamber (~ $10^{-11}$ Torr) with constant-current scanning mode. The STM tips were obtained by chemical etching from a wire of Pt(80%) Ir(20%) alloys. Lateral dimensions observed in the STM images were all calibrated using a standard graphene lattice and a Si (111)-(7×7) lattice and Ag (111) surface. The dI/dV measurements were taken with a standard lock-in technique by turning off the feedback circuit and using 793-Hz, 5 mV and 0.5 mV a.c. modulation of the sample voltage for 77 K, 4.2 K and 0.4 K measurements, respectively.

**Acknowledgements**

This work was supported by the National Natural Science Foundation of China (Grant Nos. 11674029, 11422430, 11374035, 51522212, 51421002, 51672307), the National Basic Research Program of China (Grants Nos. 2014CB920903, 2013CBA01603, 2014CB921002), the program for New Century Excellent Talents in University of the Ministry of Education of China (Grant No. NCET-13-0054), The Strategic Priority Research Program of Chinese Academy of Sciences (Grant No. XDB07030200), The Key Research Program of Frontier Sciences, CAS (Grant No. QYZDB-SSW-JSC035). L.H. also acknowledges support from the National Program for Support of Top-notch Young Professionals.

**Author contributions**

J.B.Q., W.J.Z., H.Y., N.Y, and K.Y.Q performed CVD growth of graphene/α-$Mo_2C$ heterostructures; J.B.Q. and W.J.Z. performed the main characterizations of the heterostructures, including optical, XRD, XPS, AFM, SEM and STM/STS measurements, and analyzed the data; Y.G. and J.B.Q. performed TEM measurements and analyses under the supervision of L.G. and L.H.; J.-A.S. prepared the cross section TEM sample using FIB. L.H. conceived and supervised the projects, and provided advice on the experiment and analyses. L.H., J.B.Q. and Y.G. wrote the manuscript. All the authors participated in the experiment and data discussion.

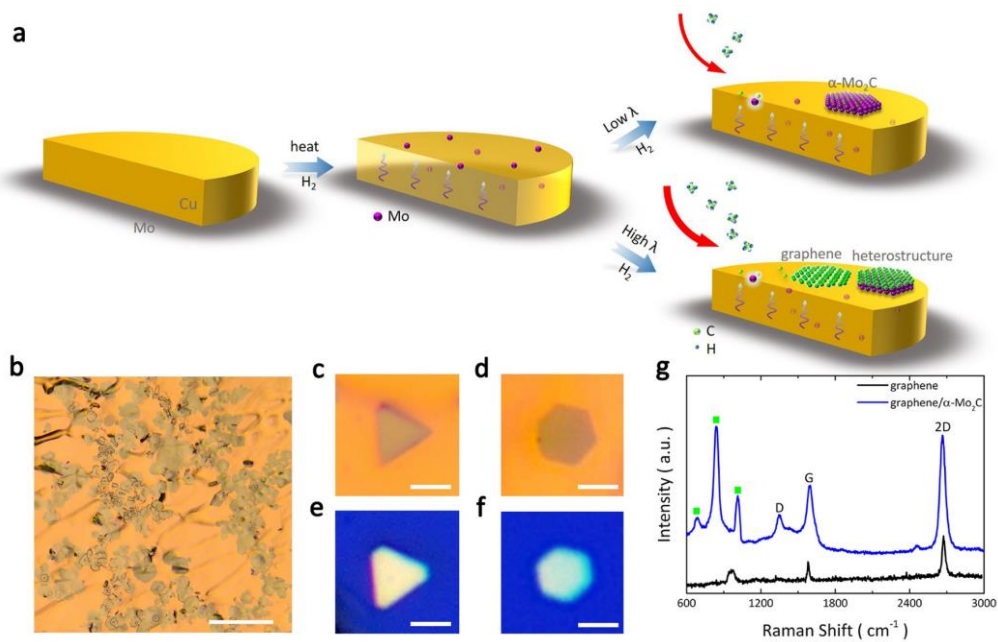

**Figure 1 | Schematic of synthesis and optical characterizations of the graphene/α-Mo$_2$C heterostructures. a**, Schematic illustration of the growth process for the heterostructures. A Cu foil is supported by a Mo substrate, and then Mo atoms start to diffuse across the liquid Cu at high temperature with the flow of H$_2$. With low ratio of CH$_4$ to H$_2$, C atoms are predominantly captured by Mo atoms (marked by luminous Mo and C atoms) and react into α-Mo$_2$C crystals. With high ratio of CH$_4$ to H$_2$, we directly obtained the graphene/α-Mo$_2$C heterostructures. **b**, Optical image of graphene/α-Mo$_2$C heterostructures on a Cu (Cu-Mo alloy)/Mo substrate. The scale bar, 50 μm. Optical images exhibit as-grown (**c** and **d**) and transferred (**e** and **f**) regular polygon-shaped heterostructures on Cu and SiO$_2$/Si targets, respectively. All the scale bars are 5 μm. **g**, Raman spectra for the transferred graphene sheets and graphene/α-Mo$_2$C heterostructures on SiO$_2$/Si targets. The green squares denote the Raman signals of α-Mo$_2$C, while the peak at low wavenumber (~ 970 cm$^{-1}$) in graphene spectrum arises from SiO$_2$/Si substrate.

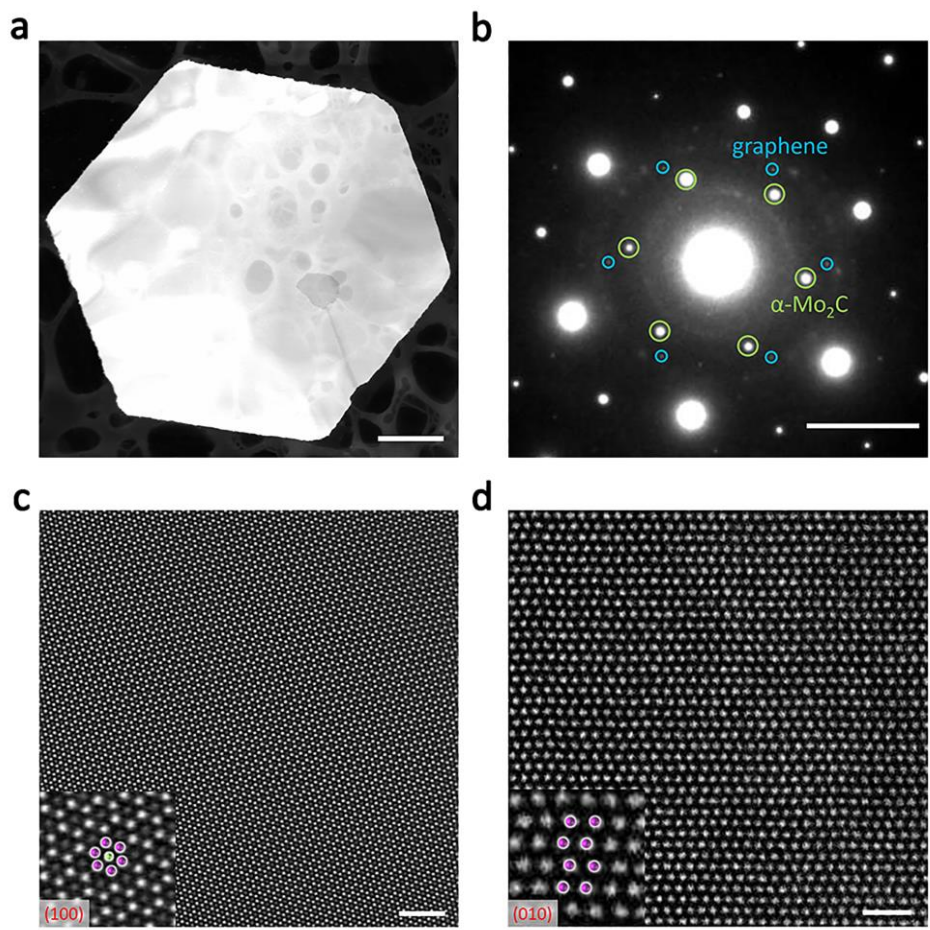

**Figure 2 | Structural characterization of the graphene/α-Mo₂C heterostructures.**
**a**, Low-magnification HAADF-STEM image of the hexagonal graphene/α-Mo₂C heterostructures. The scale bar, 1 μm. **b**, SAED pattern along the [100] zone axis, where some blue (green) spots belonging to graphene (α-Mo₂C) are circled. The scale bar, 5 nm$^{-1}$. The twist angle between graphene and α-Mo₂C is extracted as ~ 11°. Atomic-resolved HAADF-STEM images of the heterostructure surface parallel to (100) facet (**c**) and cross section parallel to (010) facet (**d**). This confirm the nature of the underlying α-Mo₂C crystals, where the center C atom (gray dot, marked by the green ball) is surrounded by six Mo atoms (white dots, marked by the purple balls) in hcp arrangement in (100) face and the Mo layers are arranged in ABAB-stacking form in (010) face (insets in **c** and **d**), respectively. Both the scale bars are 1 nm.

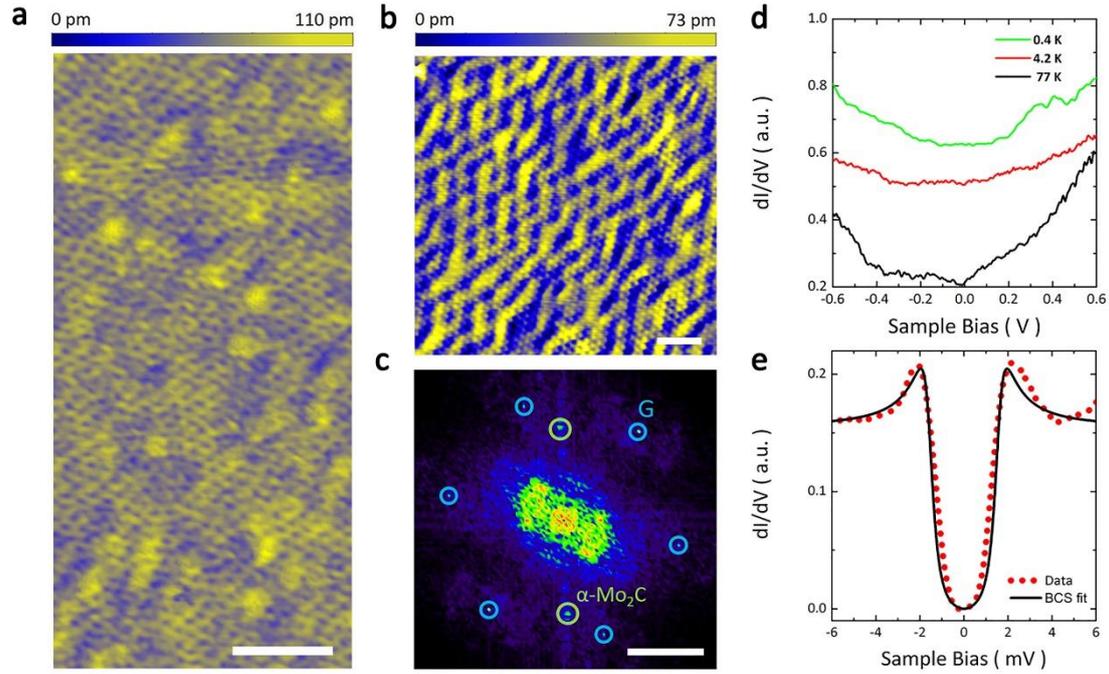

**Figure 3 | Surface topography and superconducting gap of the graphene/α-Mo₂C heterostructures. a**, Large-scale STM image of the heterostructures ($V_b$ = 1 V, $I_S$ = 0.1 nA). The scale bar, 5 nm. **b**, Topographic close-up of the patterned surface ($V_b$ = 500 mV, $I_S$ = 0.1 nA). The scale bar, 1 nm. **c**, Fast Fourier transform (FFT) image of the topography in panel **b**. The outer six spots (marked by blue circles) and an inner pair of spots (marked by green circles) are the reciprocal lattice of graphene and α-Mo₂C, respectively. The twist angle between graphene and α-Mo₂C is measured as ~ 14 °. The scale bar, 3 nm$^{-1}$. **d**, STS spectra recorded in the patterned regions of the heterostructures at different temperatures. **e**, A zooming-in dI/dV spectrum taken in the same region at 400 mK showing the superconducting gap (indicated by red dots). The black solid line is the theoretical BCS fitting with Δ = 1.58 meV and Γ = 0.4 meV.

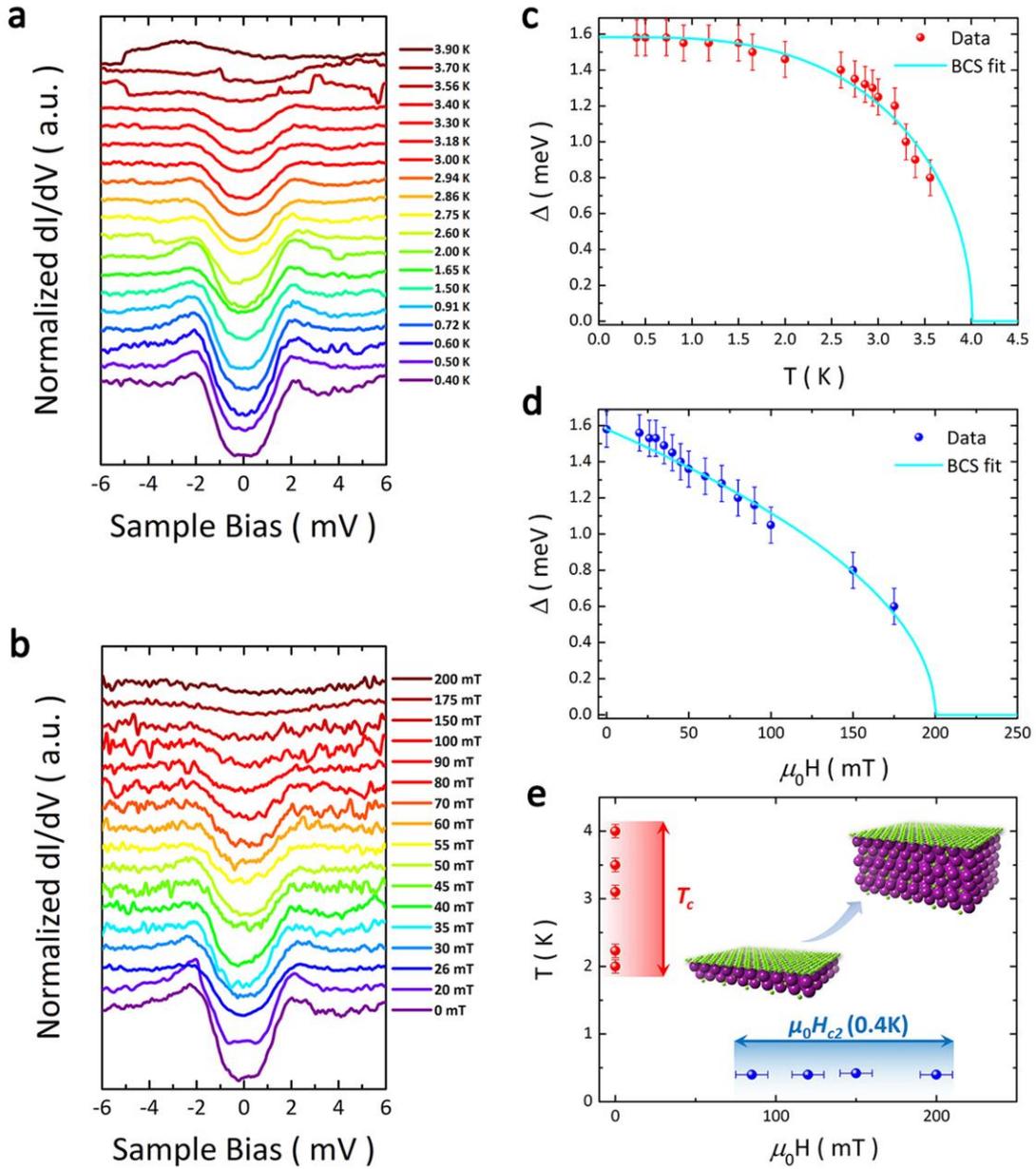

**Figure 4 | Superconducting properties of the heterostructures. a**, Temperature evolution of superconducting tunneling spectra. The curves are offset in Y-axis for clarity. a.u., arbitrary units. **b**, Normalized tunneling spectra as a function of applied magnetic fields. The curves are also offset in Y-axis for clarity. **c**, Temperature dependence of the superconducting gap. The measured Δ (T) is determined from BCS fit to the tunneling spectra, and the light blue line is theoretical result using the BCS gap function. **d**, Magnetic field dependence of superconducting gap. The measured Δ

(T) is deduced from fitting the spectra in panel **b** to the BCS density of states, consistent with the relationship, $\Delta H/\Delta 0 = \sqrt{1-H/H_c}$. **e,** Different superconducting transition temperatures and magnetic fields determined from the STS measurements. Red dots denote the variable $T_c$ in the range of 2 to 4 K, while green and blue dots indicate the critical fields $\mu_0 H_{c2}$ (T), where $\mu_0 H_{c2}$ (0.4K) varies from around 80 to 200 mT.